# Realization of type-II double-zero-index photonic crystals

Zebin Zhu[1,2†], Dong Zhao[3†], Ziyao Wang[3], Xucheng Yang[3], Liyong Jiang[1,2]*, Zhen Gao[3]*

[1]Institute of Micro-nano Photonics and Quantum Manipulation, School of Physics, Nanjing University of Science and Technology; Nanjing 210094, China.

[2]MIIT Key Laboratory of Semiconductor Microstructure and Quantum Sensing, Nanjing University of Science and Technology, Nanjing 210094, China.

[3]State Key Laboratory of Optical Fiber and Cable Manufacturing Technology, Department of Electronic and Electrical Engineering, Guangdong Key Laboratory of Integrated Optoelectronics Intellisense, Southern University of Science and Technology, Shenzhen 518055, China

†These authors contributed equally to this work.

*Corresponding author. Email: jly@njust.edu.cn (L.J.); gaoz@sustech.edu.cn (Z.G.)

**Some photonic crystals (PCs) with Dirac-like conical dispersions exhibit the property of double zero refractive index (that is, both epsilon and mu near zero (EMNZ)), wherein the electromagnetic waves have an infinite effective wavelength and do not experience any spatial phase change. The Dirac-like cones that support EMNZ are previously thought to present only at the center of the Brillouin zone (Γ point) with a zero wavevector (we refer to as type-I EMNZ), which is constrained by the proportional relationship between phase refractive index and wavevector ($n = kc/\omega$). Here, we demonstrate the existence of an anomalous type-II EMNZ in PCs, which is associated with the Dirac-like point at off-Γ points. By introducing a wave modulation approach, we theoretically elucidate its physical mechanism, and resolve the paradox of type-II EMNZ with non-zero wavevectors. We then fabricate a type-II EMNZ PC operating at the X point, and experimentally demonstrate that both its effective permittivity and permeability are zero at the Dirac-like point. Type-II EMNZ PCs exhibit a range of intriguing phenomena, including angle-selective transmission, wavefront flattening, a 180° phase shift upon transmission, and waveguiding with natural zero radiation loss. The extraordinary properties of type-II EMNZ PCs may open new avenues for the development of angle-selective optical filters, directional light sources, phase-controlled optical switches, ultracompact photonic circuits, nanolasers, and on-chip nonlinear enhancement.**

Zero index materials (ZIMs) are a special class of media in which the electromagnetic (EM) wave has an infinite wavelength and the spatial phase distribution is uniform[1-4]. This unique characteristic brings about a multitude of fascinating phenomena, such as super-coupling[5], control of emission[6,7], optical cloaking[8-10], non-reciprocal transmission[11,12], EM percolation[13], EM ideal fluid flow[14,15], and phase-match-relaxed nonlinear generation[16]. According to $n = (\varepsilon\mu)^{-1}$, ZIMs can be categorized into epsilon-near-zero (ENZ), mu-near-zero (MNZ), and epsilon-and-mu-near-zero (EMNZ) types. Among them, EMNZ media exhibit a finite non-zero impedance, enabling impedance matching with conventional materials[17]. This feature extends their utility to more promising applications, surpassing the capabilities of ENZ and MNZ media[18-21]. Howerve, it is difficult to find a material in nature that can achieve EMNZ. Fortunately, by controlling the electric/magnetic resonances, or through photonic doping of two-dimensional (2D) ENZ media, effective EMNZ has been achieved in metallic metamaterials[22], dielectric photonic crystals (PCs)[23], and waveguides at cut-off frequency[24]. Of which, EMNZ PCs have attracted special attention due to their absence of Ohmic losses[25-29] and their opportunities in three-dimensional (3D) applications[30,31].

The effective zero index in EMNZ PCs is a macroscopic property. In other words, although the local phase within a unit cell may be variable, the fields at relatively large scales behave as they would in an ideal ZIM. The EMNZ effect in PCs is often associated with an accidental degenerate Dirac-like point in momentum space, and the bands that correspond to EMNZ are typically linear[32-34]. When an EM wave propagates in an EMNZ PC, only two modes can be excited, respectively contributing to zero effective permittivity ($\varepsilon_{eff}$) and zero effective permeability ($\mu_{eff}$)[32,35,36]. To date, effective EMNZ transport has been experimentally observed at various Dirac-like points at the Brillouin zone center (Γ point) in 2D[11,23,37] and 3D[30,31] PCs.

Constrained by the proportional relationship between phase refractive index ($n_p$) and wavevector (*k*): $n_p = ck/\omega$, the Dirac-like points that can support EMNZ effect were previously thought to present only at the Γ point[23]. We refer to these as type-I EMNZ, to distinguish them from the anomalous type-II EMNZ, where Dirac-like points occur at off-Γ points with $k \neq 0$[38]. In this study, we examine a type-II EMNZ PC with a Dirac-like point at the X point. By introducing a wave modulation explanation, the physical mechanism of type-II EMNZ is elucidated, and the paradox of a zero index with $k \neq 0$ is also resolved. We fabricate such a PC sample to illustrate the existence of the Dirac-like point at X, and present direct evidence that $\varepsilon_{eff}$ and $\mu_{eff}$ are both zero at the Dirac frequency. In addition, type-II EMNZ PCs also have a range of interesting abilities, including angle-selective transmission, wavefront flattening, a 180° phase shift upon transmission, and natural zero radiation loss, which are also verified in our experiment.

**Results**

**Type-I and type-II EMNZ PCs**

Figure 1a and 1d illustrate two typical types of degenerate Dirac-like points, which are located at the Γ point and off-Γ point, corresponding to type-I and type-II EMNZ, respectively. Among them, the Dirac-like point at Γ consists of an isotropic linear Dirac-like cone with an almost flat band through it (Fig. 1a)[23]. There are three eigenmodes degenerated at the Dirac-like point in total, and two of them can be excitable by plane waves propagating along the *x*-direction (Fig. 1b) (see Supplementary Section I). It presents a superposition of an electric monopole and a transverse magnetic dipole with a 90° phase difference. When a plane wave with Dirac frequency ($f_D$) is normally incident from free space onto the PC along the *x*-direction, the instantaneous electric field distributions inside the PC oscillates synchronously over a time period of $T = 2\pi/\omega$, which is consistent with the characteristics of ideal EMNZ materials (Fig. 1c).

In contrast, due to the reduction in symmetry, the Dirac-like point at off-Γ typically manifests as a twofold-degenerate point (Fig. 1d) (see Supplementary Section I). Theoretically, the Dirac-like point of the type-II EMNZ can emerge at any location in the Brillouin zone. However, considering that the wavefront of the EM wave in free space need match the boundary of the PC, we specifically consider a Dirac-like point at X with TM polarization (the electric field is parallel to the *z* axis) to demonstrate its properties, which does not lose the generality. The Dirac-like point at off-Γ can be regarded as an isolated Dirac-like cone with an additional wavevector $k_D$. Thus, employing the inverse Fourier transform, the eigenmodes $\Psi(r)$ near the Dirac-like point can be expressed as (see Supplementary Section II):

$$\Psi(\boldsymbol{r}) = u(\boldsymbol{r})\exp(\mathrm{i}\boldsymbol{k}_\mathrm{D}\cdot\boldsymbol{r}) \qquad (1)$$

where $\boldsymbol{k}_D$ is the wavevector of the Dirac-like point, and $u(\boldsymbol{r})$ can be expanded as a superposition of a collection of plane waves according to the Bloch's theory (see Supplementary Section II):

$$u(\boldsymbol{r}) = A_0(\boldsymbol{k}_\mathrm{D} + \boldsymbol{k}')\mathrm{e}^{\mathrm{i}\boldsymbol{k}'\cdot\boldsymbol{r}} + \sum_{\boldsymbol{G}\neq 0} A_{\boldsymbol{G}}(\boldsymbol{k}_\mathrm{D} + \boldsymbol{k}')\mathrm{e}^{\mathrm{i}(\boldsymbol{k}'-\boldsymbol{G})\cdot\boldsymbol{r}} \qquad (2)$$

where $\boldsymbol{G}$ is the reciprocal lattice vector, $A_{\boldsymbol{G}}$ is the amplitude of the plane waves, and $\boldsymbol{k}' = \boldsymbol{k} - \boldsymbol{k}_\mathrm{D}$. In equation (1), we consider the high-frequency term, $\exp(\mathrm{i}\boldsymbol{k}_\mathrm{D}\cdot\boldsymbol{r})$, as the "carrier wave" generated by the lattice itself and $u(\boldsymbol{r})$ as the "modulating wave". The total field can be seen as a double sideband suppressed carrier (DSB-SC) modulation of $\exp(\mathrm{i}\boldsymbol{k}_\mathrm{D}\cdot\boldsymbol{r})$ by $u(\boldsymbol{r})$.

At the Dirac frequency, the eigenmodes (Fig. 1e) that excitable by plane waves propagating along the *x*-direction (see Supplementary Section I) does not correspond to clear electric or magnetic multipoles. And the propagating field distribution (Fig. 1f) inside the PC appears to a sinusoidal wave-like manner. Intuitively, the zero-refraction characteristics are not immediately evident. Nonetheless, after demodulating the original field, it is intriguing that both the modes (Fig. 1g) and instantaneous field distributions (Fig. 1h) exhibit the characteristics of EMNZ transport.

From a macroscopic perspective, when $|k'| \ll |k_D|$ in Equation (2), we can obtain the effective index in the direction $\hat{k}'$ as (see Supplementary Section II):

$$n_{\text{eff}} \approx \frac{ck'}{\omega} \tag{3}$$

Equation (3) shows that the effective index of the EMNZ PC is determined by $k'$, rather than $k$. When the light frequency precisely equals the Dirac frequency, the wavevector $k'$ vanish, resulting in an effective EMNZ effect.

It is notable that $u(r)$ has full translational symmetry, i.e., $u(r + \tau) = u(r)$ where $\tau$ is an arbitrary lattice vector, which ensures its value at any two unit cells are the same and supports light transport without phase delay. However, the eigenmodes of any two cells [i.e. $\Psi(r + \tau)$ and $\Psi(r)$] are generally different, unless $k_D \cdot \tau = 2\pi N$, where $N$ is an integer. In the case of the Dirac-like point at X, $\Psi(r)$ obeys an anti-symmetry under the translation operation of $(x, y) \rightarrow (x + a, y)$, i.e., $\Psi(r + a\hat{i}) = \exp(ik_D \cdot a\hat{i})\Psi(r) = -\Psi(r)$, which causes its value at $(x, y)$ and $(-x + a, y)$ between two adjacent unit cells become opposite. This signifies that the lattice number along the propagation direction must be even to achieve zero phase shift upon transmission. If the lattice number is odd, the phase of the transmitted wave is opposite to that of the incident wave, resulting in a 180° phase shift upon transmission.

**Experimental demonstration of a type-II EMNZ PC**

To verify type-II EMNZ in PCs, we fabricate a square lattice PC composed of alumina rods with twofold-degenerate bands at the X point. The lattice constant ($a$) of the PC is 22.6 mm, and the diameter ($d$) and height ($h$) of the rods are 10.45 mm and 12.35 mm, respectively. To equate the rods array to a 2D PC and facilitate TM-polarized light propagation, it is tightly sandwiched between two parallel aluminum plates. The dielectric constant of alumina is measured to be 9.77. However, since we use a PVC sticker with markers for rod positioning and use double-sided adhesive tapes to fix the rods, the dielectric constant of the rods is set to be an effective value of 9.26 when simplifying the actual PC structure to a 2D one (see "Methods" and Supplementary Section VII). The corresponding degeneracy point frequency ($f_D$) of this PC is 9.85 GHz.

First, the band structure of the PC is measured. As illustrated in Fig. 2a, a PC containing 17×17 alumina rods is excited by a dipole source at the location marked by a cyan star. The electric dipole moment direction is perpendicular to the PC. Another dipole probe antenna is passed through a series of small holes in the aluminum plate one by one into the PC to map the complex electric field distribution ($E_z$ component). Recognizing that an increase in the number of unit cells in real space can improve the resolution of a Brillouin zone, we place the source at a corner of the PC to obtain a quarter of field distribution. Subsequently, by utilizing the spatial mirror symmetry of the lattice, we flip and extend it to cover an area four times its original size, yielding a field distribution covering an equivalent PC composed of 30×30 rods, as shown in Fig. 2b at a frequency of 9.85 GHz. After applying Fourier transform to the expand field distribution from real space to reciprocal space, we obtain the band structure in the Brillouin zone. In Fig. 2c, the measured band structure is directly compared with the numerically calculated one. It is evident that the two bands are degenerated at the X point [i.e., $k = (\pm\pi/a, 0)$ and $k = (0, \pm\pi/a)$] in the Brillouin zone, with a degeneracy point frequency of 9.85 GHz, and they are linear along the Γ-X direction.

From Fig. 2b, it can be observed that EM waves only propagate along directions parallel to the x- or y-axes at $f$ = 9.85 GHz, which is due to the fact that the iso-frequency contour at this frequency is only a point at X. Therefore, the wave vector $k$ of EM waves in the PC can only be parallel to the Γ-X direction. To verify this, the transmission of a wave at 9.85 GHz when it obliquely incident on a PC with 10 layers of unit cells is measured, and it is compared with numerical simulations, as shown in Fig. 2d. The measurement method and data processing are explained in the Supplementary Section V. The figure depicts that the angle-resolved transmission spectrum exhibits a narrow peak with a half-maximum angle range of only -5.1° to 5.1°. In fact, with an increase in the layer number of unit cells, the transmitted angle range will become narrower[39] until only normally incident wave can pass through the PC. This angular selectivity of transmitted light is one of the fundamental characteristics of EMNZ materials.

Next, the near-field electric field distribution and effective optical parameters are measured when a TM-polarized plane wave is incident on a finite-width PC. As shown in Fig. 3a, an EM wave emitted from a horn antenna with polarization along z-axis is incident onto the PC that sandwiched between two parallel aluminum plates. The horn antenna is placed sufficiently far away from the PC (over 2 meters) to ensure that the incident wave can be approximated as a plane wave when it reaches the PC. Two artificial perfect magnetic conductor (PMC) boundaries are employed at the lateral edges of the PC to suppress lateral wave leakage (see "Methods" and Supplementary Section IV). A dipole probe antenna is passed through a series of small holes in the aluminum plate to map the near-field electric field distribution. Fig. 3b displays a photograph of the fabricated experimental sample, where $N$ represents the layer number of unit cells in the direction of wave propagation. For $N = 6$ or 7, Fig. 3c,d provide a comparison between the electric field distributions that simulated and measured in PCs, and the electric field distributions in ideal EMNZ materials. The simulation and measurement fields in PCs are in good agreement, both inside the PCs and in the surrounding air. By comparing the PCs with ideal EMNZ materials, the transmitted field of the PC is in phase with that of the ideal EMNZ material for $N = 6$, while they are out of phase for $N = 7$. This consistent with the theoretical prediction of the 180° phase shift upon transmission for odd-layer PCs.

For the case of the PC with $N = 6$, we extract the $S$-parameters from the measured electric field distributions of the income and outgoing waves, and then use the $S$-parameter inversion method to determine the effective permittivity ($\varepsilon_{eff}$), effective permeability ($\mu_{eff}$), and effective index ($n_{eff}$) of the PC (see "Methods" and Supplementary Section VI). Fig. 3e,f present the measured and simulated $\varepsilon_{eff}$, $\mu_{eff}$, and $n_{eff}$, and they match well with each other. At a frequency around 9.87 GHz, the curves for Re($\varepsilon_{eff}$) and Re($\mu_{eff}$) intersect at zero, and the value of Re($n_{eff}$) is also zero. This strongly supports the realization of effective EMNZ in the PC with a Dirac-like point at X. At frequencies deviating from 9.87 GHz, the small non-zero values of Im($n_{eff}$) arise due to the partial absorption of the waves with transverse momentum by artificial PMC boundaries. In Fig. 3f, the theoretical value (black solid line) of the effective index based on equation (3) is also presented, which agrees well with the simulated and measured values of Re($n_{eff}$). When the frequency of light is lower than the Dirac-like point frequency, a negative effective index is caused by $k_x' < 0$, even though $k_x$ is still positive. This results in a negative phase velocity of EM waves. Conversely, when the frequency is higher than the Dirac-like point frequency, $k_x' > 0$ leads to a positive effective index, causing a positive phase velocity of EM waves (see Supplementary Fig. S8).

**Phase flattening of arbitrary wavefronts**

One promising application of the type-II EMNZ PC is phase flattening. Here, we verify this capability by observing the electric field distribution after an EM wave with a complex wavefront passes through a 2D type-II EMNZ PC. In the experiment, a PC consisting of 8×20 alumina rods, with the same structural parameters as shown in Fig. 2, is placed between two parallel aluminum plates. A microwave signal with a frequency of 9.85 GHz is divided into five channels of equal amplitude and phase through a power divider, which are then connected to five randomly located dipoles at one side of the PC to emit a complex light field. By near field measurement using a probe antenna, the electric field distribution at the opposite side of the PC can be mapped. As a reference, we first examine the field profile without the PC. In Fig. 4a,b, respectively the simulated and measured $E_z$ field distributions directly emitted by the sources are plotted respectively, where yellow stars denote the dipole positions. The measurement is performed in a rectangular region of 200×300 mm$^2$, which is highlighted by a red dashed frame in Fig. 4a. The numerical and experimental results coincide well, revealing a quite cluttered radiation field due to interference between the sources. Next, the same light field is incident to a type-II EMNZ PC, and the simulated and measured field distributions are illustrated in Fig. 4c,d, respectively. The measured result corresponds to the blue dashed frame in Fig. 4d, showing good agreement with each other. In this case, the EM wave transmitted through the PC has a flat wavefront, with high directivity perpendicular to the PC surface. Notably, due to the excitation of zero-index modes, the type-II EMNZ PC can flatten EM waves with arbitrary wavefronts. Furthermore, the absence of interference from flat bands or other modes enables operation of wavefront flattening precisely at the Dirac frequency, resulting in the beam quality of the transmitted wave as high as possible.

**Lossless zero-index waveguides**

Another possible application of the type-II EMNZ PC is lossless zero-index waveguide. In the case of the type-I EMNZ, the presence of a zero wavevector causes it to operate above the light cone. Therefore, the mode energy will spontaneously leak into free space unless some mechanisms (such as the bound states in the continuum (BIC) effect[26-28] or additional total internal reflection[40]) are employed to eliminate out-of-plane radiation. In contrast, for the type-II EMNZ, the Dirac-like point can be designed below the light cone, implying that the modes around the Dirac-like point act as guided modes, and they do not interact with plane waves in free space when propagating in the PC.

As an example, Fig. 5 illustrates a conceptual design of a one-dimensional (1D) type-II EMNZ PC waveguide. It consists of alumina rods with a diameter of 10.45 mm, a height of 12.35 mm, and a lattice constant of 17.11 mm. Fig. 5a shows the TM-polarized band structure, including an inset of its geometrical structure. An accidentally degenerate Dirac-like point is found at the boundary of the first Brillouin zone (marked by a black arrow), corresponding to a frequency of 7.1004GHz. This location is below the light cone, which naturally allows for radiation lossless guided-mode propagation[41]. In Fig. 5b, the effective optical parameters of this PC waveguide are presented. It is apparent that the real parts of $\varepsilon_{eff}$ and $\mu_{eff}$ intersect at zero at the Dirac frequency, and their imaginary parts remain strictly zero over a broad frequency range, indicating a lossless zero-refraction propagation feature. To experimentally realize the waveguide, we utilize two rectangular dielectric waveguides to couple an EM wave into and out of the 1D PC. Additionally, a 3D microwave scanner is employed to obtain electric field distributions at sections labeled by red, yellow, and blue colors above the PC waveguide in Fig. 5c. The corresponding $E_z$ field distributions measured in the experiment are shown in Fig. 5d, which is consistent with the characteristics of the zero refractive transmission. Moreover, the EM wave propagates as an evanescent wave in the $z$ direction, and there is no observable attenuation of field intensity during propagation, indicating no apparent leakage radiation into free space.

**Discussion**

In this work, we theoretically elucidated the physical mechanism behind type-II EMNZ in PCs, and explained the paradox of a zero phase refractive index with $k \neq 0$. We designed a type-II EMNZ PC, and experimentally demonstrated the presence of a Dirac-like point at the X point of the Brillouin zone. We also directly measured the effective permittivity and effective permeability of this PC, validating the achievement of an effective EMNZ at the Dirac-like point frequency at off-Γ point. Type-II EMNZ PCs also exhibit a range of interesting properties, including angle-selective transmission, a 180° phase shift upon transmission, and natural zero radiation loss, some of which have been experimentally verified.

The outstanding optical responses of type-II EMNZ PCs have the potential to drive progress in various fields. For instance, their angular selectivity of transmission can be applied to enhance direction-dependent emission of molecular fluorescence, Raman scattering and quantum dot lighting, thereby increasing the efficiency of light signal reception, improving the sensitivity of molecular detection, or developing highly directional light sources. The PCs can also be integrated into radar domes to create angle-selective filters, which will enhance the directional emission and reception capabilities of radar signals, and strengthen their resistance to external interference. Exploiting the unique property of 180° phase shift upon transmission, type-II EMNZ PCs with an odd or even lattice number can be utilized to design phase-controlled light switches or optimize phase modulator designs. Additionally, the advantages of zero out-of-plane radiation loss enable the miniaturization and integration of the type-II EMNZ using 1D PCs, leading to wide developments in various on-chip applications, including ultracompact photonic circuits, optical interconnects, nanolasers, and on-chip nonlinear enhancement.

**Methods**

**Sample preparation.** In the cases of the 2D PCs, an array of alumina rods is tightly sandwiched between two parallel aluminum plates to create an effective 2D PC, allowing for the propagation of TM-polarized light. The designed square lattice PC has a lattice constant ($a$) of 22.6 mm, and the diameter ($d$) and height ($h$) of the rods are 10.45 mm and 12.35 mm, respectively. The dielectric constant of alumina is measured to be 9.77 at 10 GHz. To position the rods, a PVC sticker with markers is adhered to the aluminum plates. And then each rod is affixed to the corresponding marked location on the PVC sticker using double-sided adhesive tapes. The total thickness of the PVC film and adhesive tapes under the rods is approximately 0.15 mm.

In the case of the 1D PC waveguide (Fig. 5), ten alumina rods with $d = 10.45$ mm and $h = 12.35$ mm are arranged within a period of 17.11 mm. At both ends of the alumina rod array, two same alumina rectangular waveguides with a cross-section of $10.45 \times 12.35$ mm$^2$ are employed to couple the EM wave into and out of the 1D PC. To fix the positions of the 1D PC and the two rectangular waveguides, we adopt the fluted dielectric foam (ROHACELL 31 HF) with relative permittivity 1.04 and loss tangent 0.0025.

In Fig. 3a, the artificial PMC is fabricated using a double-layer FR-4 printed circuit board (PCB). The PCB has a thickness of 2 mm and a copper foil thickness of 35 μm. The top copper layer is patterned into a metasurface composed of square units with a side length of 3.4 mm and a period of 4.1 mm. The entire bottom layer is covered with copper continuously. In each unit of the metasurface, the top and bottom copper layers are connected by a via at the center of the square.

**Measurements.** We employed a vector network analyzer (VNA, Keysight E5080B) to measure the amplitude and phase of the near-field electric field. In the experiments of the 2D PCs, the microwave antennas are connected to the output port of the VNA to emit the EM waves. A dipole probe, which connected to the input port of the VNA, is passed through a series of small holes in the aluminum plate one by one into the PC to map the electric field distributions. In experiment of the 1D PC waveguide, the dipole probe is fixed on a 3D scanning movement platform to detect the electric near field distributions.

**Numerical simulation.** All numerical results presented in this work are simulated using the RF module of COMSOL Multiphysics 6.0. The band structure is simulated by a 2D approach. Due to the presence of PVC sticker and double-sided tapes under the alumina rods, an effective dielectric constant of 9.26 is adopted for the cylinders in 2D simulation. The transmission properties of the PC are simulated by a 3D approach, employing the actual dielectric constant of materials. We also take into account of the PVC sticker, adhesive tapes, and manufacturing errors to make the simulation results as close to the experimental data as possible.

**Data availability**

Data supporting key conclusions of this work are included within the article and Supplementary information. All data that support the plots within this paper and other findings of this study are available from the corresponding author upon reasonable request.

**Code availability**

The code that supports the plots within this paper and other findings of this study is available from the corresponding authors upon reasonable request.

**Acknowledgments**

L.J. acknowledges support from the National Natural Science Foundation of China under Grants No. 61675096, Six Talent Climax Foundation of Jiangsu under Grants No. XYDXX-027, Fundamental Research Funds for the Central Universities under Grants No. 30922010801, Fundamental Research Funds for NUST under Grants No. TSXK2022D006. The work at Southern University of Science and Technology was sponsored by the National Natural Science Foundation of China (No. 62375118, No 6231101016, and No. 12104211), Guangdong Basic and Applied Basic Research Foundation (No. 2024A1515012770), and Shenzhen Science and Technology Innovation Commission (No. 20220815111105001).

**Authors Contributions**

Z.Z. and L.J. initiated the idea. L.J. supervised the project, with consultation with Z.G. Z.Z. carried out the analytical derivations and full-wave simulations. Z.Z. and D.Z. designed the experiments. D.Z., X.Y., and Z.W. fabricated samples and performed experiment measurements. Z.Z. and D.Z. analyzed data. Z.Z. wrote the manuscript with inputs from D.Z. and Z.G. L.J. and Z.G. revise the manuscript.

**Competing Interests**

The authors declare no competing interests.

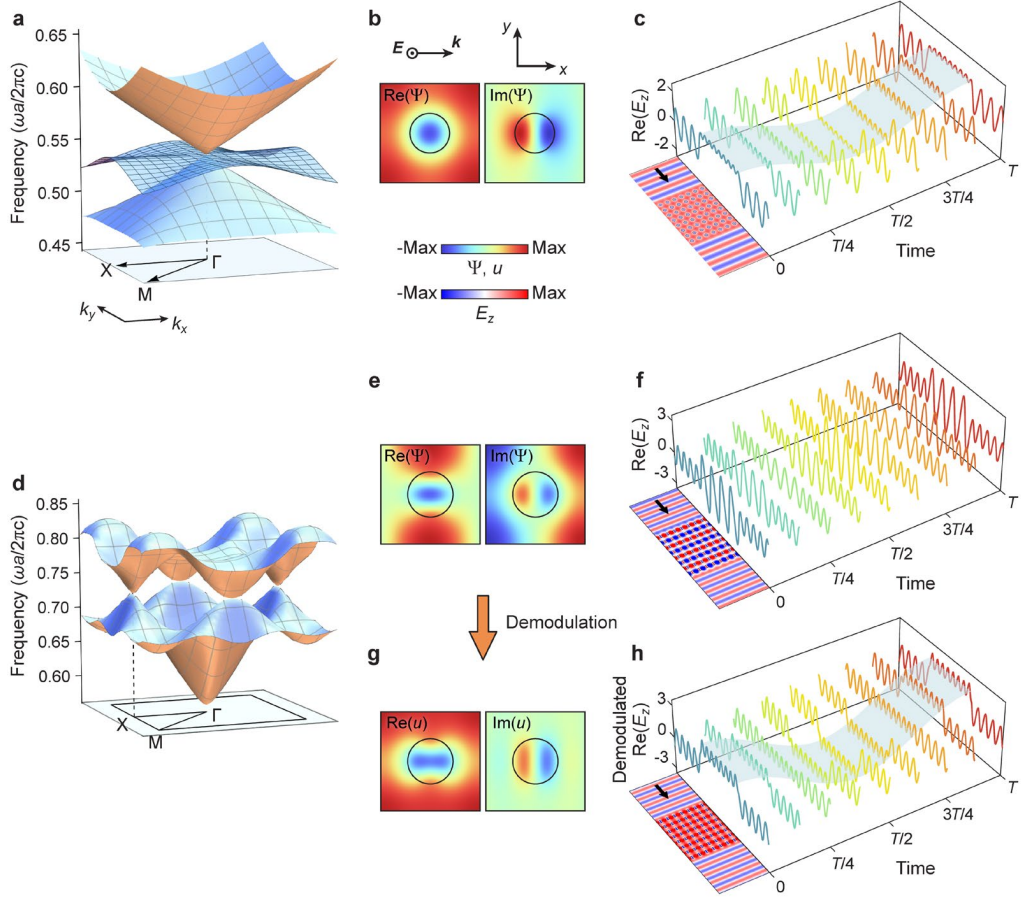

**Fig. 1 | A comparison of the type-I and type-II effective EMNZ in PCs. a**, **d** TM-polarized dispersion near the Dirac-like points at Γ and X, with dielectric constants of the rods being 12.5 (**a**) and 9.77 (**d**), and radii being 0.2$a$ (**a**) and 0.23$a$ (**d**). The degeneracy point frequencies $f_D$ are 0.5413c/$a$ (**a**) and 0.7344c/$a$ (**d**), respectively. **b**, **e** Eigenmodes Ψ($r$) that are excitable by plane waves with a group velocity along the $x$-direction at the degeneracy points. **c**, **f** Instantaneous electric field distributions in the propagation direction at different times when a plane wave with the frequency of $f_D$ is normally incident onto the PCs, where $T = 2\pi/\omega$ is the time period of the EM waves. The left sides of the 3D boxes depict the overall electric field distributions at $t = 0$. **g** The mode $u(r)$ obtained after demodulating the eigenmodes (**e**) at the Dirac-like point at X. **h** Instantaneous field distributions after demodulating the original electric field (**f**) inside the PC over one time period.

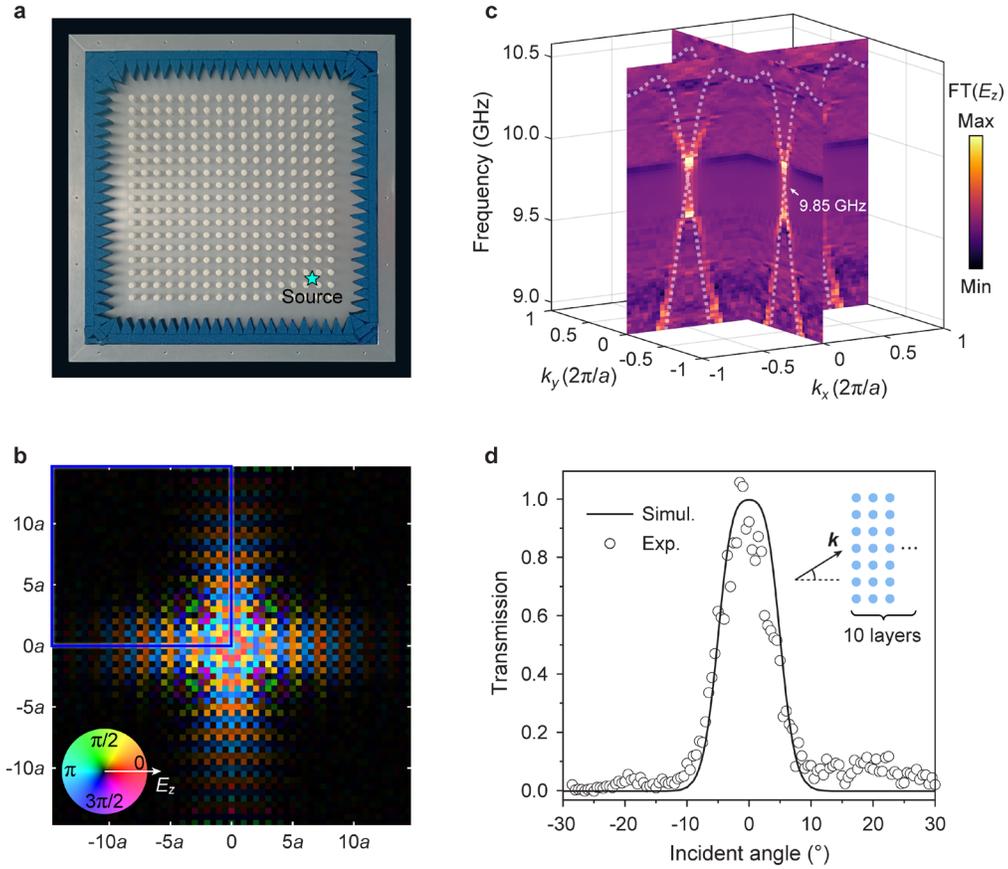

**Fig. 2 | Experimental validation of the twofold-degenerate Dirac-like point at the X point in a PC. a** Photograph of the fabricated PC that consists of 17×17 alumina rods. The dipole source (cyan star) is placed at a corner of the PC. **b** The electric field distribution ($E_z$) in real space scanned at 9.85 GHz, where the brightness and color represent the amplitude and phase of $E_z$, respectively. The blue square highlights the directly measured field, which is then flipped to obtain the field in the other three quadrants due to the spatial mirror symmetry of the lattice. **c** Measured (background color) and calculated (white dashed lines) band structures along the $\Gamma$-$k_x$ and $\Gamma$-$k_y$ directions, with a degeneracy point at the frequency of 9.85 GHz. **d** Comparison of the measured and simulated angle-resolved transmission spectrum at 9.85 GHz when a plane wave obliquely incident on a PC with 10 layers of unit cells.

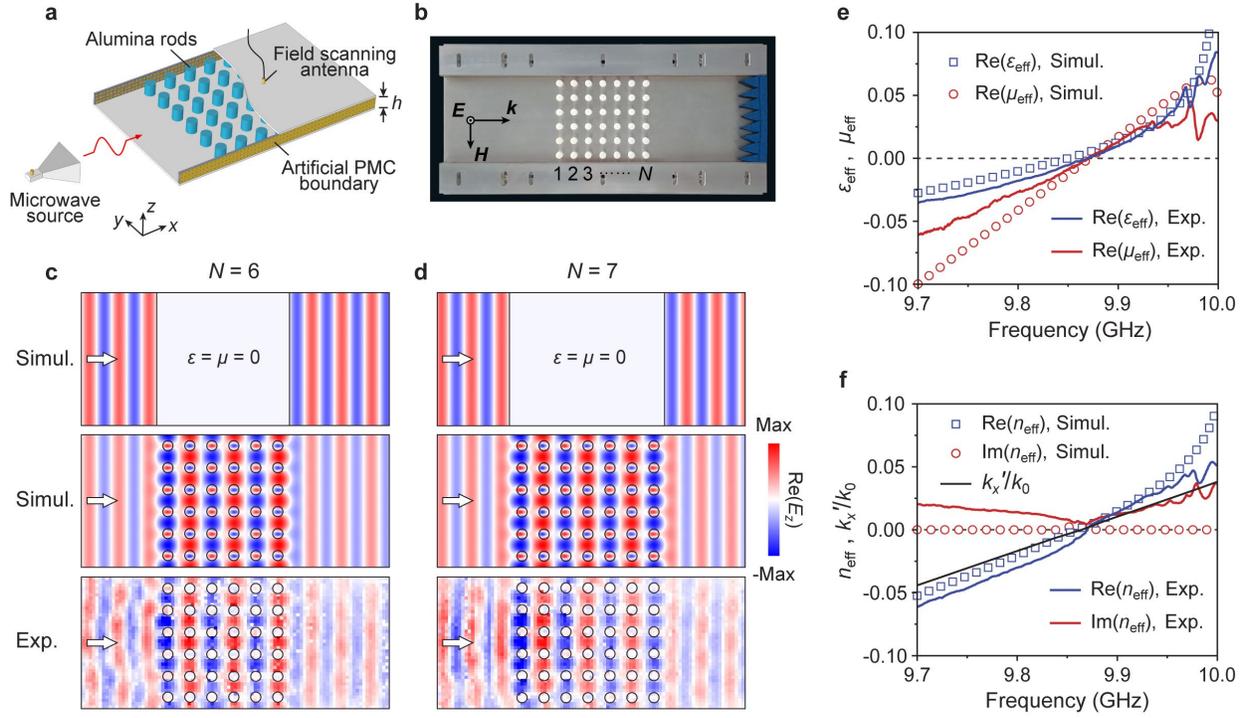

**Fig. 3 | Experimental verification of the type-II EMNZ in a PC. a**, **b** Schematic diagram and photograph of the experimental setup used to scan the near-field electric field distributions and inverse the effective optical parameters, where *N* represents the layer number of the unit cells in the direction of wave propagation. **c**, **d** Separately for *N* = 6 or 7, the simulated near-field electric field distribution when a plane wave with a frequency of 9.87 GHz propagates through an ideal EMNZ material, as well as the simulated and measured results when the same plane waves propagate through PCs. **e**, **f** Simulated and measured results of the effective permittivity, effective permeability, and effective index versus frequency. The black solid line in f is the result calculated from equation (3).

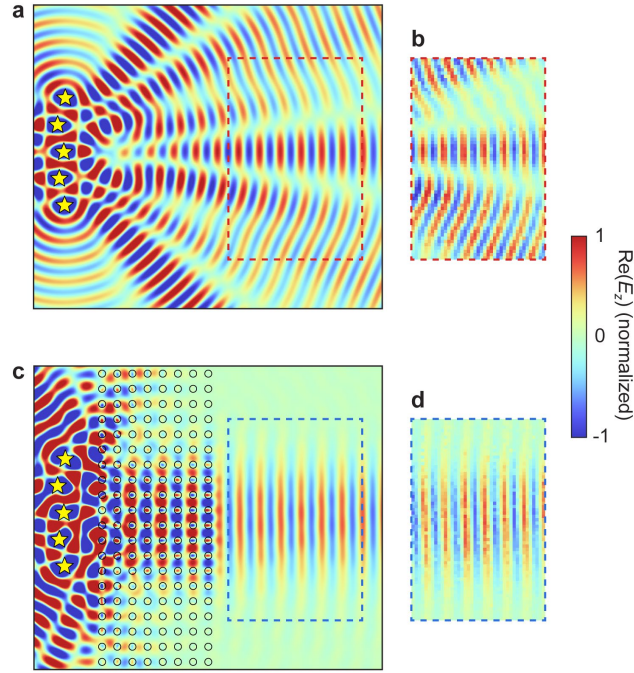

**Fig. 4 | Demonstration of wavefront flattening by a type-II EMNZ PC. a,c** Simulated electric field distributions of EM waves emitted by five randomly placed dipole sources (marked by yellow stars) with (**a**) and without (**c**) the type-II EMNZ PC at 9.85 GHz. The red and blue dashed frames in (**a**) and (**c**) depict the measured regions in the experiment, respectively. **b,d** The measured field distributions corresponding to (**a**) and (**c**), respectively.

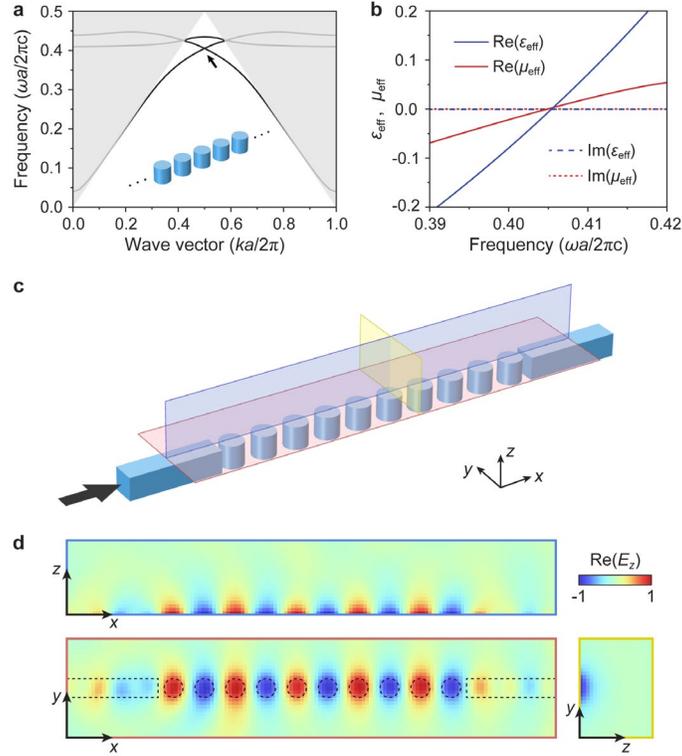

**Fig. 5 | Design of a lossless zero-index waveguide.** The one-dimensional (1D) PC is composed of alumina rods with a diameter of 10.45 mm, a height of 12.35 mm, and a lattice constant of 17.11 mm. **a** TM-polarized band structure of the 1D PC, where the shaded area denotes the light cone region, and the black arrow marks a Dirac-like point. The inset depicts the geometrical structure of the PC waveguide. **b** Simulated effective optical parameters of the PC waveguide. **c** Geometric sketch of a PC waveguide utilized for experimental measurements, with red, yellow, and blue colored sections representing the near-field scanning planes in the experiment. **d** Measured electric field distribution at 7.1004 GHz corresponding to the three sections in (**c**).

# Supplementary Information

# Realization of type-II double-zero-index photonic crystals

Zebin Zhu[1,2†], Dong Zhao[3†], Ziyao Wang[3], Xucheng Yang[3], Liyong Jiang[1,2]*, Zhen Gao[3]*

[1]Institute of Micro-nano Photonics and Quantum Manipulation, School of Physics, Nanjing University of Science and Technology; Nanjing 210094, China.

[2]MIIT Key Laboratory of Semiconductor Microstructure and Quantum Sensing, Nanjing University of Science and Technology, Nanjing 210094, China.

[3]State Key Laboratory of Optical Fiber and Cable Manufacturing Technology, Department of Electronic and Electrical Engineering, Guangdong Key Laboratory of Integrated Optoelectronics Intellisense, Southern University of Science and Technology, Shenzhen 518055, China

†These authors contributed equally to this work.

*Corresponding author. Email: jly@njust.edu.cn (L.J.); gaoz@sustech.edu.cn (Z.G.)

# I. k·p perturbation and coupling between eigenmodes

## A. k·p perturbation of the eigenmodes

In the coupled bands with a total of $N$ bands that contain the Dirac-like point, let $\Psi_{n,k}(r)$ represents the mode of the $n$-th band at a wavevector $k$, where $n = 1, 2, 3 \ldots, N$. This mode can be expanded as a superposition of the perturbations of all the eigenmodes $\Psi_{j,k_D}(r)$ at the Dirac-like point ($k_D$), i.e.

$$\Psi_{n,k}(r) = \sum_j \psi_{nj}(k) e^{i(k-k_D)\cdot r} \Psi_{j,k_D}(r) \tag{S1}$$

Here, $|\psi_n(k)\rangle = (\psi_{n1}, \psi_{n2}, \cdots \psi_{nN})^T$ represents the proportionality coefficient, and it satisfies an eigenfunction:

$$\hat{H}|\psi_n(k)\rangle = \frac{\omega_n^2(k)}{c^2}|\psi_n(k)\rangle \tag{S2}$$

According to Ref. [1], the Hamiltonian in equation (S2) is given by:

$$H_{lj} = \frac{\omega_j^2(k_D)}{c^2}\delta_{lj} - (k-k_D)\cdot p_{lj} + (k-k_D)^2 q_{lj} \tag{S3}$$

with

$$p_{lj} = i\frac{(2\pi)^2}{\Omega}\iint_{\text{unitcell}} \Psi^*_{l,k_D}(r)\left\{\frac{2\nabla\Psi_{j,k_D}(r)}{\mu_r(r)} + \left[\nabla\frac{1}{\mu_r(r)}\right]\Psi_{j,k_D}(r)\right\}d^2r \tag{S4}$$

$$q_{lj} = \frac{(2\pi)^2}{\Omega}\iint_{\text{unitcell}} \Psi^*_{l,k_D}(r)\frac{\Psi_{j,k_D}(r)}{\mu_r(r)}d^2r \tag{S5}$$

## B. The propagation mode at the Dirac-like point at Γ

There are three eigenmodes corresponding to Fig. 1a in the main text; they are electric monopole ($\Psi_{1,k_D}(r)$), transverse magnetic dipole ($\Psi_{2,k_D}(r)$), and longitudinal magnetic dipole ($\Psi_{3,k_D}(r)$), as shown in Fig. S1a. It is notable that the eigenmodes are normalized, i.e.

$$\Psi(r) = \frac{E_z(r)}{\sqrt{\langle E_z(r)|\varepsilon_r(r)|E_z(r)\rangle}}$$
$$= E_z(r)\left[\frac{(2\pi)^2}{\Omega}\iint_{\text{unitcell}} E_z^*(r)\varepsilon_r(r)E_z(r)dr^2\right]^{-\frac{1}{2}} \tag{S6}$$

Substituting the eigenmodes into the eigenfunction (S2), the eigenvalues and the eigenstates can be solved. The eigenvalues that vary with $k$ form three bands, marked as Upper,

Middle, and Lower branches. The cross-section of them along the Γ-X direction, as well as the simulated counterparts, are depicted in Fig. S1b. When an electromagnetic (EM) wave propagates inside the PC in the x-direction, the activated bands are the branches with group velocity ($\partial\omega/\partial k_x$) being positive, i.e. the Upper branch with $k_x > 0$ or the Lower branch with $k_x < 0$, depending on the wave frequency. Thus, at the Dirac frequency, the propagation mode is the mode at $k_x = 0^+$ of the Upper branch, or the mode at $k_x = 0^-$ of the Lower branch.

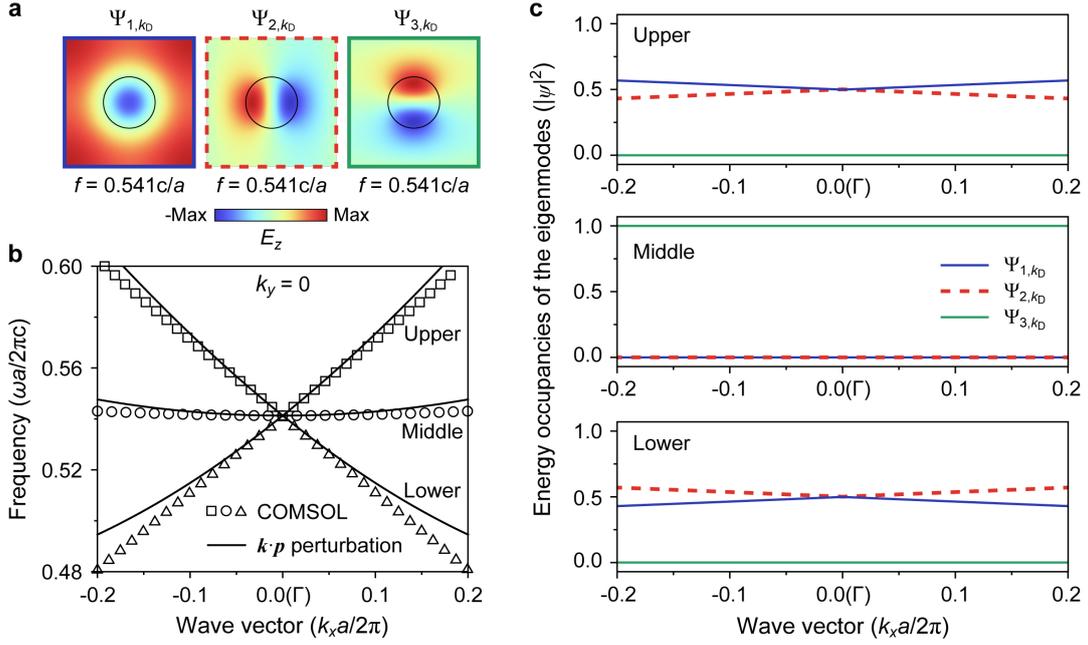

**Fig. S1 | $k \cdot p$ perturbation near the Dirac-like point at Γ.** The structure parameters of the PC are the same as those in Fig. 1a of the main text. **a** Three degenerate eigenmodes at the Dirac-like point. **b** The band structure along the Γ-X direction ($k_y = 0$) near the Dirac-like point, where the square, circle and triangular markers illustrate the simulated bands using COMSOL, while the solid lines depict the calculated counterpart via the $k \cdot p$ perturbation. **c** The energy occupancies of each eigenmode in Upper, Middle, and Lower branches obtained via the $k \cdot p$ perturbation.

Fig. S1c illustrates the energy occupancies of the three modes in each branch. The propagation mode of the $n$-th band at the Dirac-like point is a linear combination of $\Psi_{j,\boldsymbol{k}_\mathrm{D}}$:

$$\Psi_{n,\boldsymbol{k}_\mathrm{D}} = \psi_{n1}\Psi_{1,\boldsymbol{k}_\mathrm{D}} + \psi_{n2}\Psi_{2,\boldsymbol{k}_\mathrm{D}} + \psi_{n3}\Psi_{3,\boldsymbol{k}_\mathrm{D}} \tag{S7}$$

where $n$ represents Upper, Middle, or Lower. From Fig. S1c, it can be observed that both the Upper or Lower branches contain two modes of $\Psi_{1,\boldsymbol{k}_\mathrm{D}}$ and $\Psi_{2,\boldsymbol{k}_\mathrm{D}}$, and the energy contributions of them are both 50% at Γ. While the Middle branch solely contains the mode $\Psi_{3,\boldsymbol{k}_\mathrm{D}}$. It is solved that the eigenstate at $k_x = 0^+$ of the Upper branch and at $k_x = 0^-$ of the Lower branch are both $\left|\psi_{\text{Upper}}(\boldsymbol{k}_D^+)\right\rangle = \left|\psi_{\text{Lower}}(\boldsymbol{k}_D^-)\right\rangle = \left(1/\sqrt{2},\ i/\sqrt{2},\ 0\right)^T$, indicating their equivalence. Consequently, at the Dirac-like point, the propagation mode in the *x*-direction is given by:

$$\Psi_{\text{propagation}} = \frac{1}{\sqrt{2}}\Psi_{1,k_D} + i\frac{1}{\sqrt{2}}\Psi_{2,k_D} \tag{S8}$$

This corresponds to the mode presented in Fig. 1b of the main text.

## C. The propagation mode at the Dirac-like point at X

The Dirac-like point at X, which proposed in this work, is degenerated from two bands (TM-4 and TM-5 in Fig. S2b). However, the two adjacent bands (TM-3 and TM-6 in Fig. S2b) will interact with them. Hence, when studying the $\boldsymbol{k}\cdot\boldsymbol{p}$ perturbation near the Dirac-like point at the X-point, it is necessary to consider four eigenmodes [2], as shown in Fig. S2a. Substituting the four eigenmodes into the eigenfunction (S2), the eigenvalues at different wavevectors $\boldsymbol{k}$ can be solved, constituting four branches. The dispersion relationships along the Γ-X and X-M directions of the four branches are depicted in Fig. S2b,c, respectively. It is notable that the DZI effect in this PC occurs only when waves propagate in the Γ-X direction. Therefore, only the modes in this direction need consideration when we study the DZI effect. In Fig. S2b, when an EM wave with a frequency near $f_D$ propagates inside the PC in the *x*-direction, the activated bands are the branches with positive group velocity ($\partial\omega/\partial k_x$), i.e. the TM-5 branch with $k_x > \pi/a$, or the TM-4 branch with $k_x < \pi/a$, depending on the wave frequency. Thus, at the Dirac frequency, the propagation mode is the mode at the right side of X ($X^+$) in the TM-5 branch, or the mode at the left side of X ($X^-$) in the TM-4 branch.

Fig. S2d,e present the energy occupancies of each eigenmode in the four branches along the orthogonal Γ-X and X-M directions. In the Γ-X direction (Fig. S2d), TM-4 and TM-5 bands contain only $\Psi_{2,k_D}$ and $\Psi_{3,k_D}$, and both the two modes occupy 50% of the energy. While $\Psi_{1,k_D}$ and $\Psi_{4,k_D}$ make no contributions for TM-4 and TM-5 bands. Hence, the generation of DZI effect solely relates to the modes $\Psi_{2,k_D}$ and $\Psi_{3,k_D}$. It is solved that the eigenstate at $X^+$ of the TM-5 branch and at $X^-$ of the TM-4 branch are both $\left|\psi_{\text{TM-5}}(\boldsymbol{k}_D^+)\right\rangle = \left|\psi_{\text{TM-4}}(\boldsymbol{k}_D^-)\right\rangle = \left(0,\ -i/\sqrt{2},\ 1/\sqrt{2},\ 0\right)^T$, indicating their equivalence. Consequently, at the Dirac-like point, the propagation mode in the *x*-direction is given by:

$$\Psi_{\text{propagation}} = \frac{1}{\sqrt{2}}\Psi_{3,k_D} - i\frac{1}{\sqrt{2}}\Psi_{2,k_D} \tag{S9}$$

This corresponds to the mode presented in Fig. 1e of the main text.

In the X-M direction (Fig. S2e), the TM-4 band is affected by the mode $\Psi_{4,k_D}$ at non-

Dirac point, and the TM-5 band is affected by the mode $\Psi_{1,k_D}$ at non-Dirac point. This signify that the adjacent TM-3 and TM-6 bands will interact with the Dirac-like cone at the direction along X-M.

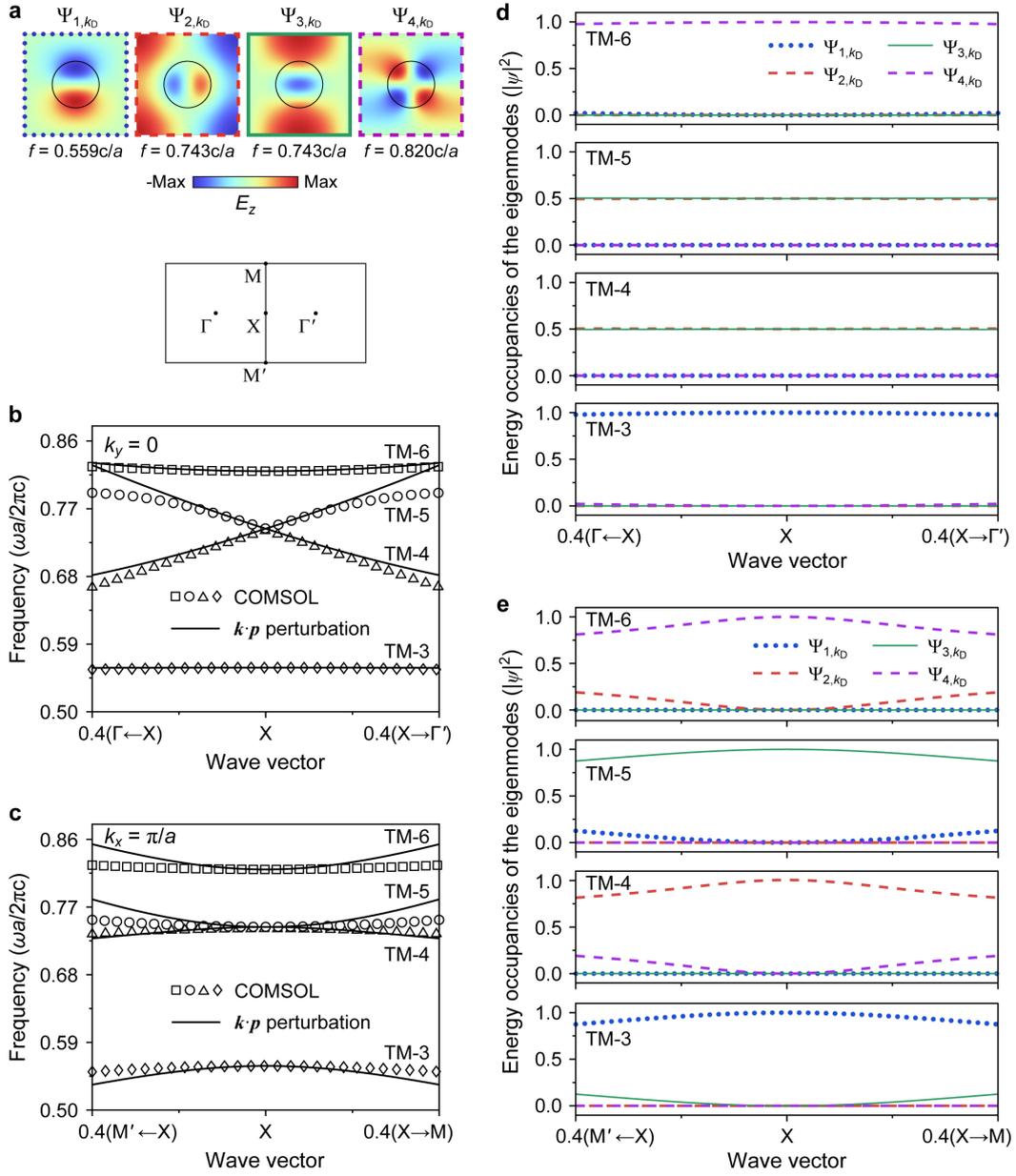

**Fig. S2 | $k \cdot p$ perturbation near the Dirac-like point at X.** The structure parameters of the PC are the same as those in Fig. 1d of the main text. **a** Four eigenmodes at the Dirac-like point (($k_x$, $k_y$) = ($\pi/a$, 0)) and the adjacent two bands. **b**, **c** The band structure along the Γ-X ($k_y = 0$) and X-M ($k_x = \pi/a$) directions near the Dirac-like point, where the square, circle and triangular markers illustrate the simulated bands using COMSOL, while the solid lines depict the calculated counterpart via the $k \cdot p$ perturbation. **d**, **e** Corresponding to **b** and **c**, the energy occupancies of each eigenmode in four branches obtained via the $k \cdot p$ perturbation.

## II. Effective phase delay of the modulating wave

According to the Bloch's theory, the eigenmodes of photonic crystals (PCs) can be expanded as:

$$\Psi_k(r) = u_k(r) e^{ik \cdot r} \tag{S10}$$

where

$$u_k(r) = \sum_G A_G(k) e^{-iG \cdot r} \tag{S11}$$

Here $G$ is the reciprocal lattice vector given by $G = l_1 b_1 + l_2 b_2$, where $b_1$ and $b_2$ are base vectors in reciprocal space, and $l_1$ and $l_2$ are both integers. Let $k = k_D + k'$, we get:

$$\Psi_k(r) = u_{k_D}(r) e^{ik_D \cdot r} \tag{S12}$$

where

$$\begin{aligned} u_{k_D}(r) &= u_k(r) e^{ik' \cdot r} \\ &= \sum_G A_G(k_D + k') e^{i(k'-G) \cdot r} \\ &= A_0(k_D + k') e^{ik' \cdot r} + \sum_{G \neq 0} A_G(k_D + k') e^{i(k'-G) \cdot r} \end{aligned} \tag{S13}$$

Equation (S12) and (S13) correspond to equation (1) and (2) in the main text, respectively. When $|k'| \ll 2\pi/a$, and if $G \neq 0$, it follows that $|k'| \ll |G|$. Therefore, equation (S13) can be approximately written as:

$$u_{k_D}(r) \approx A_0 e^{ik' \cdot r} + \sum_{G \neq 0} A_G e^{-iG \cdot r} \tag{S14}$$

Taking the average within a unit cell, we get:

$$\langle u_{k_D}(r) \rangle_{\text{unitcell}} \approx \frac{A_0}{\Omega} \iint_{\text{unitcell}} e^{ik' \cdot r} d^2 r + \frac{1}{\Omega} \sum_{G \neq 0} A_G \iint_{\text{unitcell}} e^{-iG \cdot r} d^2 r \tag{S15}$$

where $\Omega = \iint_{\text{unitcell}} 1 \cdot d^2 r$ is the area of a unit cell. For two-dimensional (2D) PCs, denoting $a_1$ and $a_2$ as base vectors in real space, an arbitrary coordinate in a unit cell can be expressed as $r = x_1 a_1 + x_2 a_2$, where $x_1, x_2 \in [0, 1]$. Noting that $G = l_1 b_1 + l_2 b_2$, the integral in the second term of equation (S15) can be written as:

$$\iint_{\text{unitcell}} e^{-iG \cdot r} d^2 r = \iint_{\text{unitcell}} \exp[-i(l_1 b_1 + l_2 b_2) \cdot (x_1 a_1 + x_2 a_2)] dx_1 dx_2 \tag{S16}$$

Using $a_i \cdot b_j = \delta_{ij}$, where $\delta_{ij}$ is the Kroenke symbol, equation (S16) can be further derived as:

$$= \iint_{\text{unitcell}} e^{-i2\pi(x_1 l_1 + x_2 l_2)} dx_1 dx_2 = \int_0^1 e^{-i2\pi x_2 l_2} \left[ \int_0^1 e^{-i2\pi x_1 l_1} dx_1 \right] dx_2 \tag{S17}$$

Since $l_1$ and $l_2$ are both integers, the integral vanishes. Thus equation (S15) can be simplified by:

$$\left\langle u_{k_D}(r) \right\rangle_{\text{unitcell}} \approx \frac{A_0}{\Omega} \iint_{\text{unitcell}} e^{ik'\cdot r} \, d^2r \quad (S18)$$

This is equation (3) in the main text. We can observe from equation (S18) that the effective phase delay of the collective mode oscillation of $u_{k_D}(r)$ between adjacent unit cells depends solely on the low-frequency component $A_0\exp(ik'\cdot r)$.

## III. Point group symmetry of the modes

In this section, we will determine the irreducible representations of the modes $\Psi_{k_D}(r)$ and $u_{k_D}(r)$.

The point group at the X point of a square lattice has a $C_{2v}$ symmetry, which contains four symmetry operations: the identity operation $\hat{E}$, a 180° rotation $\hat{C}_2$, a reflection along the x-axis $\hat{\sigma}_x$, and a reflection along the y-axis $\hat{\sigma}_y$. Each operation (denoted by $\hat{R}$) corresponds to a matrix representation $D_{\hat{R}}$. The relationship between the mode $f(r)$ and the operation $\hat{R}$ is defined as:

$$\hat{R}f(r) = \chi(\hat{R})f(r) \tag{S19}$$

where $\chi(\hat{R}) = \mathrm{Tr}(D_{\hat{R}})$ represents the character of the representation. The $C_{2v}$ group has four irreducible representations, denoted by $A_1$, $A_2$, $B_1$, and $B_2$, corresponding to four sets of characters, respectively, as listed in Table S1 [3].

**Table S1 | Character $\chi$ table for group $C_{2v}$**

| Irreducible representation | Symmetry operations | | | |
|---|---|---|---|---|
| | $\hat{E}$ | $\hat{C}_2$ | $\hat{\sigma}_y$ | $\hat{\sigma}_x$ |
| $A_1$ | 1 | 1 | 1 | 1 |
| $A_2$ | 1 | 1 | −1 | −1 |
| $B_1$ | 1 | −1 | 1 | −1 |
| $B_2$ | 1 | −1 | −1 | 1 |

**Table S2 | Character $\chi$ of the modes $\Psi_k(r)$ and $u_{k_D}(r)$**

| Modes | Symmetry operations | | | |
|---|---|---|---|---|
| | $\hat{E}$ | $\hat{C}_2$ | $\hat{\sigma}_y$ | $\hat{\sigma}_x$ |
| $\mathrm{Re}(\Psi_k)$ | 1.0000 | 0.9989 | 1.0000 | 0.9989 |
| $\mathrm{Im}(\Psi_k)$ | 1.0000 | −0.9993 | 1.0000 | −0.9993 |
| $\mathrm{Re}(u_{k_D})$ | 1.0000 | 0.9998 | 1.0000 | 0.9998 |
| $\mathrm{Im}(u_{k_D})$ | 1.0000 | −0.9950 | 1.0000 | −0.9950 |

The irreducible representations of the $C_{2v}$ group are all one-dimensional, which implies that modes at X are not inherently degenerate (although accidental degeneracy may occur).

Therefore, the character of a mode $f(\mathbf{r})$ can be calculated using the following equation:

$$\chi(\hat{R}) = \frac{\iint\limits_{\text{unitcell}} \varepsilon(\mathbf{r})f(\mathbf{r})\hat{R}f(\mathbf{r})\mathrm{d}^2\mathbf{r}}{\iint\limits_{\text{unitcell}} \varepsilon(\mathbf{r})f^2(\mathbf{r})\mathrm{d}^2\mathbf{r}} \quad (S20)$$

By substituting the modes $\Psi_{k_D}(\mathbf{r})$ and $u_{k_D}(\mathbf{r})$ shown in Fig. 1e,g of the main text into equation (S20), their characters are calculated as listed in Table S2. Comparing Table S1 and S2, it is evident that $\mathrm{Re}(\Psi_{k_D})$, $\mathrm{Im}(\Psi_{k_D})$, $\mathrm{Re}(u_{k_D})$, and $\mathrm{Im}(u_{k_D})$ belong to the irreducible representations $A_1$, $B_1$, $A_1$, and $B_1$, respectively. Hence, $\mathrm{Re}(\Psi_{k_D})$ and $\mathrm{Re}(u_{k_D})$ have the same point group symmetry, while $\mathrm{Im}(\Psi_{k_D})$ and $\mathrm{Im}(u_{k_D})$ have the same point group symmetry.

## IV. Artificial perfect magnetic conductor (PMC) boundary

The artificial PMC metasurfaces are fabricated using a double-layer FR-4 printed circuit board (PCB). As displayed in Fig. S3a,b, the PCB has a thickness of 2 mm with a copper foil thickness of 35 μm. The top copper layer is etched to form a metasurface composed of square shapes with a side length of 3.4 mm and a period of 4.1 mm. The entire bottom layer is covered with copper continuously. A via with an inner diameter of 0.4 mm and an outer diameter of 0.6 mm connects the top and bottom copper foils at the center of the square. Fig. S3c displays a photograph of the artificial PMC boundary we fabricated.

The dielectric constant of the board material, FR-4, is measured to be 4.238 (1 + 0.016i) at 10 GHz. We employ COMSOL software to simulate the reflectance and the associated phase shift of plane waves normally incident upon the metasurface, as shown in Fig. S3d. At a frequency of 9.9 GHz, the phase shift upon reflection is zero, consistent with the characteristic of an ideal PMC [4], with a reflectance of 0.87.

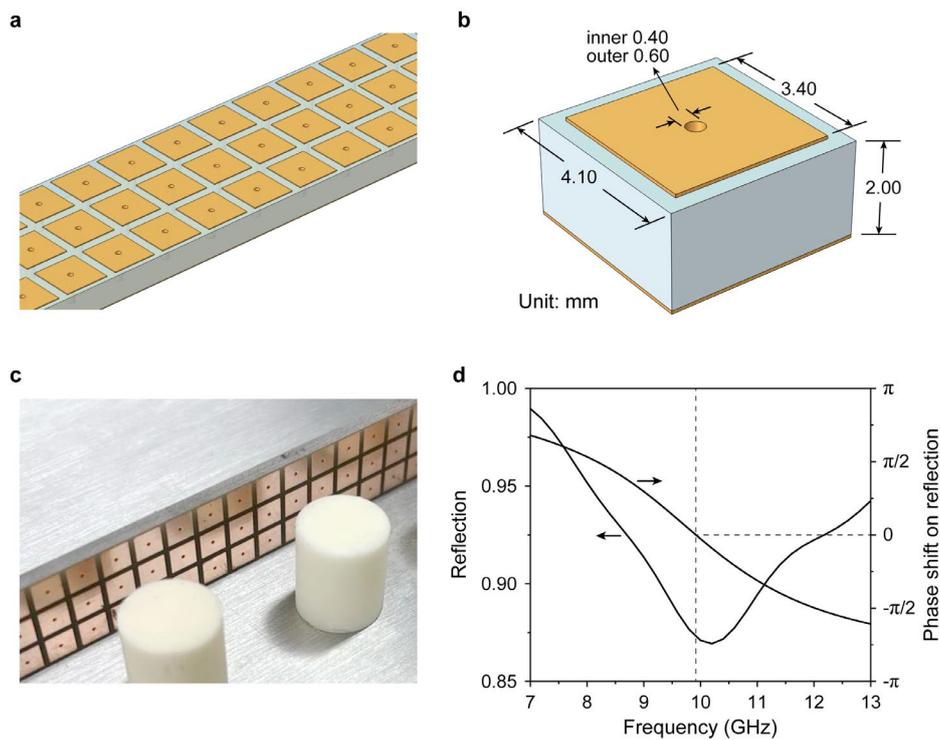

**Fig. S3 | Artificial perfect magnetic conductor (PMC) metasurface. a** Schematic diagram of the designed structure. **b** Enlarged view of a unit cell. **c** Photograph of the fabricated sample. **d** Simulated reflectance and the associated phase shift of the artificial PMC metasurface.

## V. Experiment setup of measuring angle-resolved transmission spectra

As sketched in Fig. S4, A rotating platform is constructed to measure the angular-resolved spectrum of the PC. The PC, composed of a 10×30 array of alumina dielectric rods, is sandwiched between two parallel circular aluminum plates to form an equivalent 2D PC. A horn antenna is placed at one side of the PC to emit TM-polarized EM wave. The antenna is sufficiently far from the PC (over 2 meters) to ensure that the incident wave can be approximated as a plane wave when it reaches the PC. At the other side of the PC, an arced aperture is set in the above aluminum plate, and it is coaxial with the rotator. A probe antenna is passed through the arced aperture into the parallel aluminum plate to measure the transmitted electric field intensity. The PC, the rotational axis of the rotator, and the probe antenna are all kept aligned with the optical axis of the EM waves emitted from the horn antenna.

During measurements, we firstly measure the field intensity $E_{z0}^2$ within the two parallel aluminum plates without the PC. Then, with the addition of the PC, we measure the field intensity $E_z^2(\alpha)$ in the range of -30° to 30° with a step of 0.5° by rotating the rotator. It is notable that the probe's position relative to the horn antenna should remain unchanged while the PC is in rotation. Finally, the transmittance is obtained by:

$$T(\alpha) = \frac{E_z^2(\alpha)}{E_{z0}^2} \tag{S21}$$

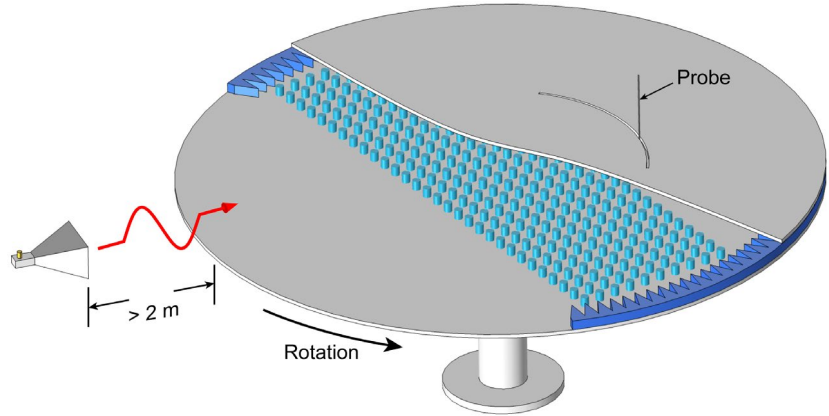

**Fig. S4 | The experimental setup for measuring the angle-resolved transmission spectra.**

## VI. Data processing method for measuring effective optical parameters

The effective optical parameters ($\varepsilon_{\text{eff}}$, $\mu_{\text{eff}}$, and $n_{\text{eff}}$) of the PC are inverted from the S-parameters, which are extracted from the measured near-field electric field distributions.

### A. Data processing of the experimental near-field electric field

Specifically, we firstly use a vector network analyzer (VNA, Keysight E5080B) to scan the complex electric field distributions $E_z(x,y,\omega)$ of the PC within a parallel metal plate waveguide, as sketched in Fig. S5a. As an example, the measured electric field distribution at 9.87 GHz is exhibited in Fig. S5b,c. In the input channel ($x < -3.5a$) and the output channel ($x > 3a$), the EM waves primarily propagate in the form of plane waves, with a small proportion of scattered waves. To extract the plane wave components ($E_{\text{p.w.}}(x,\omega)$), we perform an overlap integral between the directly measured field ($E_z(x,y,\omega)$) and the wavefront of a plane wave ($E_0(y) = 1$):

$$E_{\text{p.w.}}(x,\omega) = \frac{\int E_0^*(y) E_z(x,y,\omega) \mathrm{d}y}{\int E_0^*(y) E_0(y) \mathrm{d}y} \tag{S22}$$

The distribution of $E_{\text{p.w.}}(x)$ at 9.87 GHz obtained from experimental data, based on equation (S22), is plotted by scattered points in Fig. S5d,e.

### B. Extraction of S-parameters

Next, we fit the wave equation of the plane wave to the scattered points shown in Fig. S5d,e to obtain the S-parameters.

In the input channel ($x_{\text{edge}} < x < x_{\text{in}}$) of the structure in Fig. S5a, the incident waves ($A_1$ and $A_1'$) propagate in the positive x-direction. A part of them enter the PC, while another part ($A_2$) are reflected by the PC and propagate in the negative x-direction. When the reflected waves ($A_2'$) reach the left edge (at $x = x_{\text{edge}}$) of the aluminum plate, a part of them ($A_3$) are reflected again and propagate in the positive x-direction. The waves undergo multiple reflections in the input channel, forming standing waves. Let $S_{11}$, $r_{\text{edge}}$, and $\beta$ denote the reflection coefficient of the PC, the reflection coefficient of the edge of parallel aluminum plate, and the propagation constant within the parallel metal plate waveguide, respectively. We can establish the following equations:

$$A_j' = S_{11} A_{j+1} \text{ , for } j \text{ being odd at } x = x_{\text{in}} \tag{S23}$$

$$A_j' = r_{\text{edge}} A_{j+1} \text{ , for } j \text{ being even at } x = x_{\text{edge}} \tag{S24}$$

$$A_j' = A_j \exp[i\beta(x_{\text{in}} - x_{\text{edge}})] \tag{S25}$$

Based on equations (S23) to (S25), the total electric field at coordinate $x$ within the input channel ($x_{\text{edge}} < x < x_{\text{in}}$) is a summation of the incident field and all reflected fields, given by:

$$E_{in}(x) = (A_1 + A_3 + \ldots)e^{i\beta(x-x_{edge})} + (A_2 + A_4 + \ldots)e^{i\beta(x_{in}-x)}$$
$$= \frac{A_1}{1-e^{2i\beta(x_{in}-x_{edge})}S_{11}r_{edge}}\left[e^{i\beta(x-x_{edge})} + S_{11}e^{i\beta(2x_{in}-x_{edge}-x)}\right] \quad (S26)$$

We determine $r_{edge} = 0.182\exp(-1.682i)$ and $\beta = k_0 = \omega/c$ by simulations using COMSOL. Let $A_1$ and $S_{11}$ as parameters, we fit equation (S26) to the experimental data within $x < x_{in}$ in Fig. S5d,e, and obtain the reflection coefficient ($S_{11}$) of the PC, as depicted by the black curves in Fig. S5f,g. Additionally, the electric field at the PC's input plane ($x = x_{in}$) can be subsequently calculated by $E_{in}(x_{in})$.

Analogously, within the output channel ($x_{out} < x < x_{abs}$), after a superposition of all EM waves, the total electric field is expressed as:

$$E_{out}(x) = \frac{B_1}{1-e^{2i\beta(x_{abs}-x_{out})}r_{abs}S_{11}}\left[e^{i\beta(x-x_{out})} + r_{abs}e^{i\beta(2x_{abs}-x_{out}-x)}\right] \quad (S27)$$

Here we take into account the reflection ($r_{abs}$) of the wave-absorbing material, although it is almost zero. Let $B_1$ and $r_{abs}$ as fitting parameters, equation (S27) is fitted to the experimental data within $x > x_{out}$ in Fig. S5d,e, resulting in the output electric field $E_{out}(x_{out})$ at the PC's output plane ($x = x_{out}$). Then, as depicted by the rad curves in Fig. S5f,g, the transmission coefficient of the PC is calculated through:

$$S_{21} = \frac{E_{out}(x_{out})}{E_{in}(x_{in})} \quad (S28)$$

## C. Optical parameter inversion

After obtaining $S_{11}$ and $S_{21}$ of the PC, the effective impedance and effective index can be calculated by [5]:

$$Z_{eff} = \sqrt{\frac{(1+S_{11})^2 - S_{21}^2}{(1-S_{11})^2 - S_{21}^2}} \quad (S29)$$

$$n_{eff} = -\frac{i}{k_0(x_{out}-x_{in})}\ln\frac{S_{21}}{1-S_{11}(Z_{eff}-1)/(Z_{eff}+1)} \quad (S30)$$

Thereby the effective permittivity and effective permeability, which are illustrated in Fig. 3e,f of the main text, is obtained by:

$$\varepsilon_{eff} = n_{eff}/Z_{eff} \quad (S31)$$

$$\mu_{eff} = n_{eff}Z_{eff} \quad (S32)$$

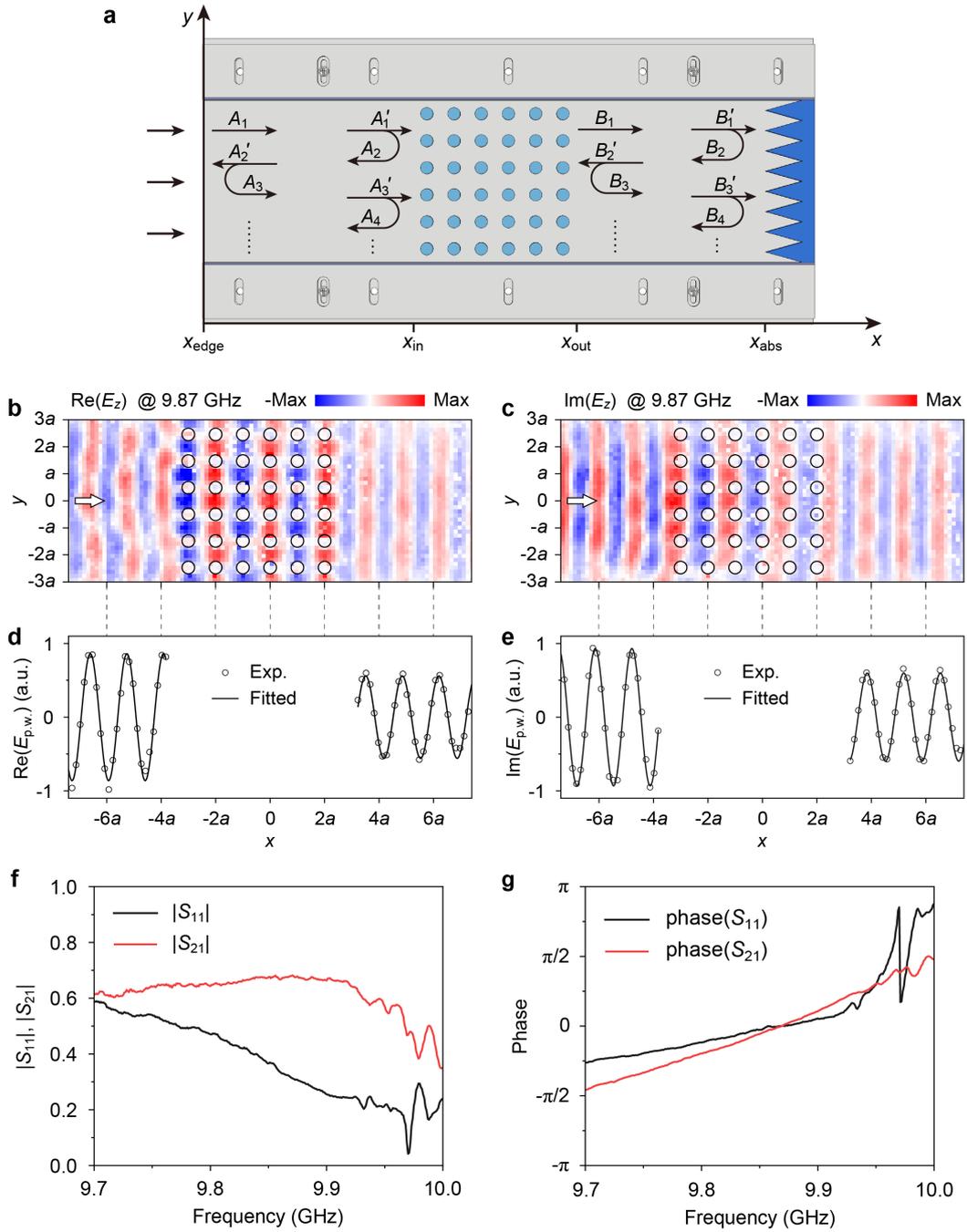

**Fig. S5 | Data processing method for measuring the effective optical parameters. a** Schematic diagram of EM waves propagating in a parallel metal plate waveguide. **b, c** Real and imaginary parts of the measured $E_z$ field distributions at 9.87 GHz. **d, e** Extracted plane wave components of the propagating fields (scatters) from **a** and **b**, and the fitted curves according to the wave functions (solid lines). **f, g** After fitting the wave function at each frequency point, the obtained magnitude and phase of $S_{11}$ and $S_{21}$ versus frequency.

## VII. Numerical simulation details

All simulations in this work are performed using the RF module of COMSOL Multiphysics 6.0. The band structure is simulated through a two-dimensional (2D) method, whereas the transmission properties are simulated through a three-dimensional (3D) method.

### A. Settings for 3D simulation

A unit cell of the PC sandwiched between two parallel aluminum plates is sketched in Fig. S6a,b. alumina rods with a lattice constant of 22.6 mm are positioned using a PVC sticker, and affixed to the aluminum plate using double-sided tapes. The diameter, height and dielectric constant of the rods are 10.45 mm, 12.35 mm and 9.77, respectively. Below the alumina rods, a total thickness of the PVC film and double-sided tape is 0.15 mm, with a dielectric constant of 2.38 [6]. Above the alumina rods, an air gap is introduced between rods and the aluminum plate to account for their rough surfaces, which causes the experimental Dirac frequency to be higher than the design value. The air gap width is chosen to be 0.04 mm to match the experiment. In addition, the manual adherence of the alumina rods onto the aluminum plate may lead to some deviations in the lattice constant.

Angular-resolved spectra, near-field distributions, and effective optical parameters of the PC (Fig. 2d, 3, and 4 of the main text) are all simulated using a 3D method, where the influences from the PVC sticker, double-sided tapes, and the air gap are considered. As shown in Fig. S6c, the boundaries in the $x$-direction are set as ports, and those in the $y$-direction are set as Floquet periodic boundary conditions (PBCs).

For the simulation of the angular-resolved spectra, the number of rods ($N$) is 10, with a lattice constant of 22.6 mm. The transmittance is calculated by $T = S_{21}^2$.

For the simulation of the near-field distributions and the effective optical parameters, the number of rods ($N$) is set to 6 or 7, and the lattice constant is adjusted to a practically measured value of 22.4 mm due to manual pasting errors. The effective optical parameters are calculated using equations (S29)-(S32) by substituting $S_{11}$ and $S_{21}$.

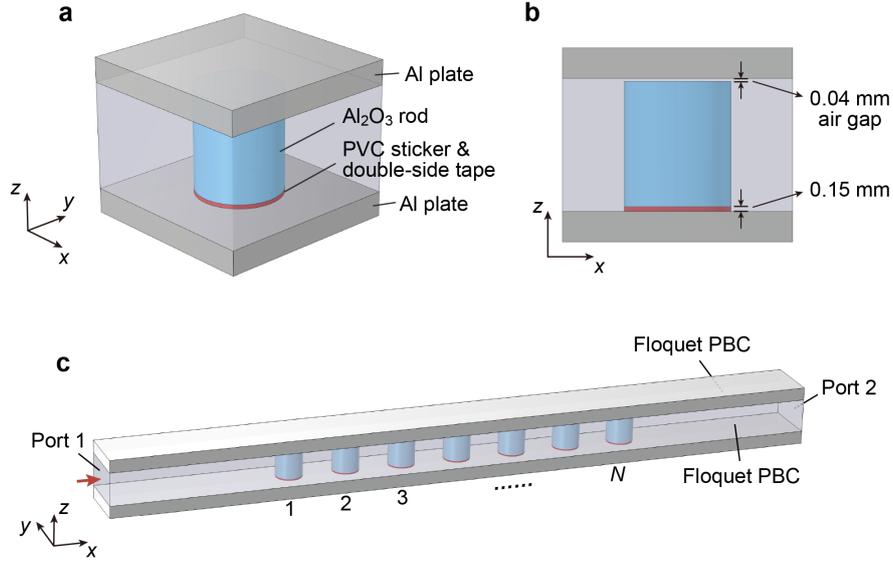

**Fig. S6 | Structure of the 3D simulation. a, b** The structure of a single unit cell. A layer of PVC positioning sticker and double-sided tape with a total thickness of 0.15 mm is below the alumina rod, and an air gap with a width of 0.04 mm above. **c** Structure for simulating the angle-resolved spectra, near-field electric field distributions, and effective optical parameters, where the boundaries in the *x*-direction are set as ports, and those in the *y*-direction are set as Floquet periodic boundary conditions (PBCs).

### B. Settings for 2D simulation

When simulating the band structure using a 2D method, we simplify the composite pillar of "PVC sticker - double-sided tape - alumina - air gap" into an infinitely high cylinder with uniform optical parameters. Thus, its dielectric constant should adopt an equivalent value $\bar{\varepsilon}_{rod}$.

To determine the equivalent dielectric constant of the composite pillar, we perform a 2D simulation [7]. As sketched in Fig. S7a, the composite structure with a width of *d* is placed in a parallel metal plate waveguide to simulate its *S*-parameters. And then the equivalent dielectric constant of it is calculated using:

$$\bar{Z}_{rod} = \sqrt{\frac{(1+S_{11})^2 - S_{21}^2}{(1-S_{11})^2 - S_{21}^2}} \tag{S33}$$

$$\bar{n}_{rod} = -\frac{i}{k_0 d} \ln \frac{S_{21}}{1 - S_{11}(\bar{Z}_{rod}-1)/(\bar{Z}_{rod}+1)} \tag{S34}$$

$$\bar{\varepsilon}_{rod} = \bar{n}_{rod} / \bar{Z}_{rod} \tag{S35}$$

Different values of *d* have a minimal impact on $\bar{\varepsilon}_{rod}$. As displayed in Fig. S7b-d, with an increase in *d*, some higher-order resonant modes may be excited inside $Al_2O_3$, causing the waves within Al$_2$O$_3$ not to be plane waves anymore. To obtain $\bar{\varepsilon}_{rod}$ as accurately as possible, we set $d$ to be a relatively small value of 1 mm, and get $\bar{\varepsilon}_{rod} = 9.26$ at 10 GHz.

The band structure (dashed line in Fig. 2c of the main text) is simulated using a 2D method. A square unit cell with a lattice constant of 22.6 mm contain a 10.45-mm-diameter cylinder, whose diameter and dielectric constant are 10.45 mm and 9.26, respectively. The boundaries are set as Floquet periodic boundary conditions (PBCs) in both the $x$- and $y$-directions. By varying the wave vector $\mathbf{k}$ of the Floquet periodicity, the PC's band structure in reciprocal space is solved.

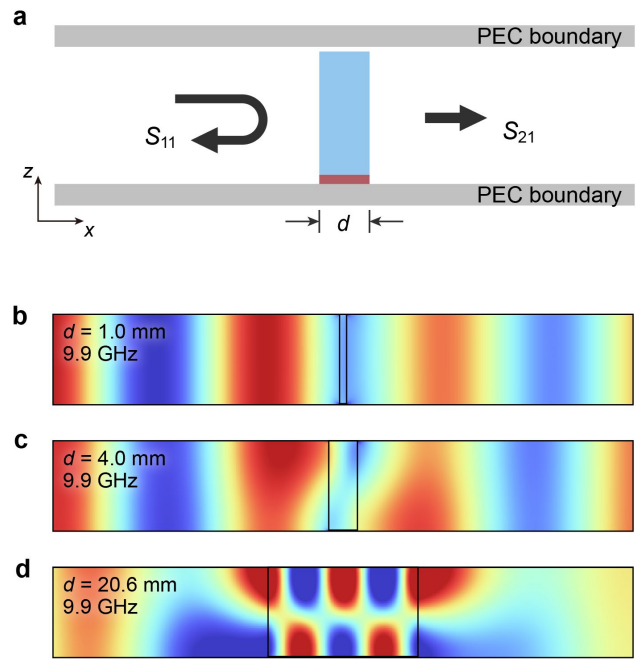

**Fig. S7 | 2D simplification of the "PVC sticker - double-sided tape - alumina - air gap" composite pillar. a** Diagram of the equivalent approach, where the aluminum plate is simplified to a perfect electric conductor (PEC) boundary. **b-d** The $E_z$ distributions of the composite structures with different widths ($d$).

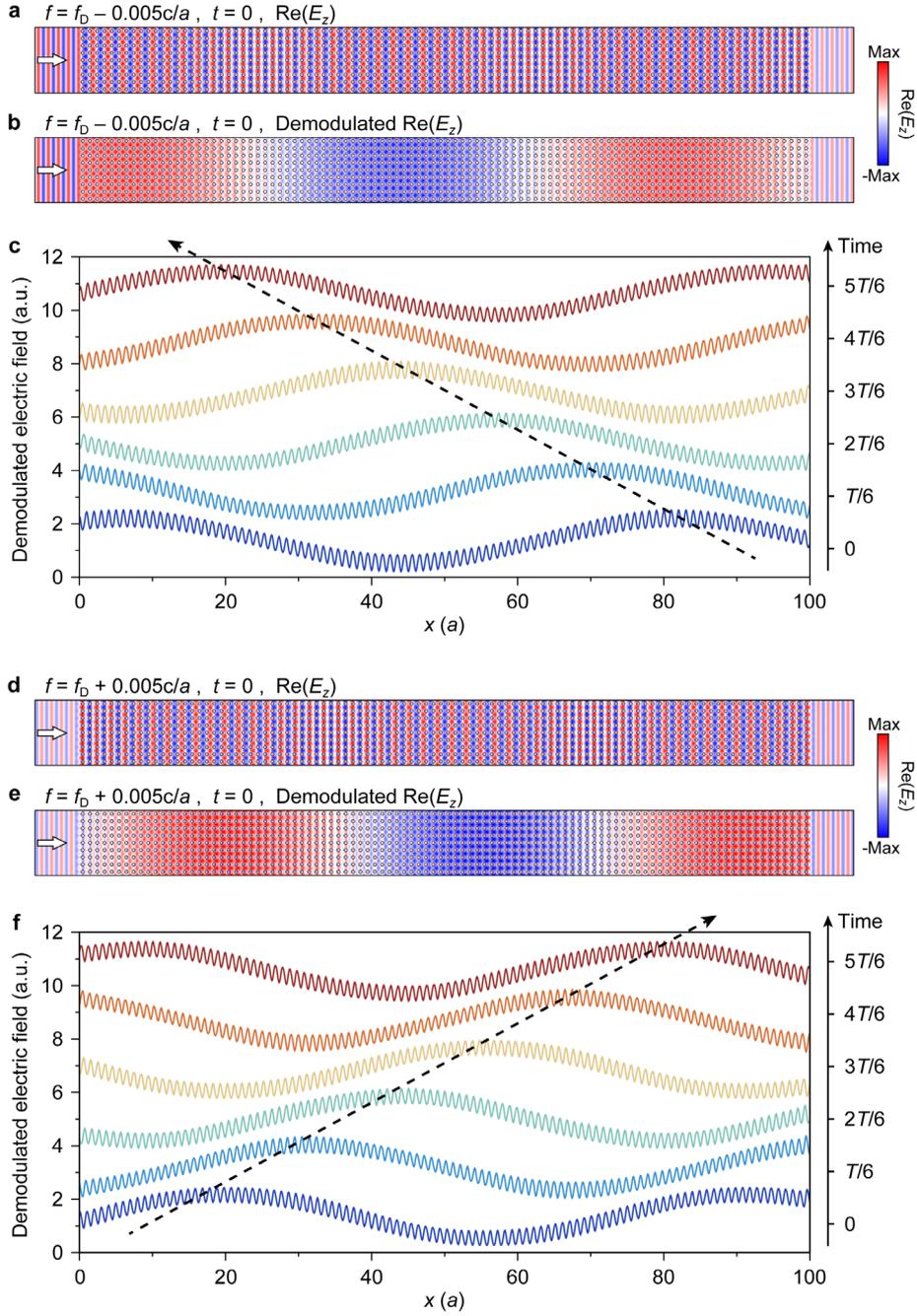

**Fig. S8 | The movement of the wavefront when frequency deviating from the Dirac frequency ($f_D$).** The structure parameters are the same as those in Fig. 1d of the main text, with a lattice number of 100 along the propagation direction and $f_D = 0.7344\ c/a$. At a frequency of $f = f_D - 0.005\ c/a$, **a** the instantaneous electric field distribution and **b** its demodulated counterpart in PC. **c** The demodulated field distributions along the propagation direction at different time, where the black dashed arrow indicates the wavefront movement, and $T = 2\pi/\omega$ is the time period. **d-f** The same as **a-c**, but $f = f_D + 0.005c/a$.

# Supplementary References